# Unbalanced synaptic inhibition can create intensity-tuned auditory cortex neurons

Andrew Y. Y. Tan[1,3], Craig A. Atencio[1], Daniel B. Polley[1,2], Michael M. Merzenich[1] and Christoph E. Schreiner[1]

[1]*Coleman Memorial Laboratory and W.M. Keck Foundation Center for Integrative Neuroscience, University of California, San Francisco, California 94143*

[2]*Present address: Vanderbilt Kennedy Center for Human Development and Department of Hearing & Speech Sciences, Vanderbilt University Medical Center, Nashville, Tennessee 37232-8548*

[3]*Email: atyy@alum.mit.edu*

## Abstract

Intensity-tuned auditory cortex neurons may be formed by intensity-tuned synaptic excitation. Synaptic inhibition has also been shown to enhance, and possibly even create intensity-tuned neurons. Here we show, using in vivo whole cell recordings in pentobarbital-anesthetized rats, that some intensity-tuned neurons are indeed created solely through disproportionally large inhibition at high intensities, without any intensity-tuned excitation. Since inhibition is essentially cortical in origin, these neurons provide examples of auditory feature-selectivity arising de novo at the cortex.

## Introduction

The intensity of a sound is often behaviourally important. For example, it can convey the distance of a sound source, or prosodic information in speech and music. Most auditory cortex neurons have monotonic spike rate versus sound intensity functions; they encode sound intensity by increasing spike rate as sound intensity is increased. A different encoding strategy, however, is used by the intensity-tuned neurons whose spike rates are nonmonotonic functions of sound intensity: their spike rate initially increases and peaks as sound intensity is increased, then decreases as sound intensity is further increased (Phillips et al 1985; Phillips and Kelly 1989). The number of intensity-tuned auditory cortex neurons increases in rats trained to perform a task requiring fine intensity discrimination, which suggests that intensity-tuned neurons are required for sound intensity to be precisely encoded (Polley et al 2004).

There are no intensity-tuned neurons at the auditory periphery, so central inhibition is required for the formation of intensity-tuned neurons. As there are intensity-tuned neurons at subcortical auditory stations, that central inhibition is not necessarily cortical (Rhode and Smith 1986; Ding and Voigt 1997; Ding et al 1999; Davis and Young 2000; Pollak et al 1978; Ryan and Miller 1978, Palombi and Caspary 1996a, 1996b; Sivaramakrishnan et al 2004).

However, Ojima and Murakami's (2002) intracellular study of the cat auditory cortex described 12 'unbalanced' intensity-tuned neurons which received disproportionally large synaptic inhibition at high intensities. As Ojima and Murakami (2002) did not measure synaptic excitation and inhibition separately, they could not determine whether inhibition enhanced intensity-tuning already present in the excitation, or whether the interaction of inhibition with excitation actually created intensity-tuning that was not present in the excitation. In Wehr and Zador's (2003) intracellular study of the rat auditory cortex synaptic excitation and inhibition were separately measured. Wehr and Zador (2003) described 7 'balanced' intensity-tuned neurons which received intensity-tuned synaptic excitation and identically tuned synaptic inhibition which neither created nor enhanced intensity-tuning. The small sample size of Wehr and Zador (2003) would not be inconsistent with unbalanced intensity-tuned neurons being present in the auditory cortex of the rat, as they are in the cat's.

We therefore performed in vivo whole cell measurements of the synaptic excitation and inhibition received by intensity-tuned auditory cortex neurons of pentobarbital-anesthetized rats. To increase the chances of obtaining in vivo whole cell recordings from intensity-tuned neurons in these rats, we increased the number of such neurons in their auditory cortices using a variant of the above mentioned training (Polley et al 2004, 2006).

This work has been previously published in abstract form (Tan et al 2005).

## Methods

### Behavioural training

All experimental procedures used in this study were approved under UCSF Animal Care Facility protocols. Adult female Sprague-Dawley rats were trained as described by Polley et al (2006). Rats were trained to identify a target auditory stimulus from a set of distracter auditory stimuli. The auditory stimuli consisted of 150 ms tone pips (10 ms linear ramps) of variable frequency and intensity. Rats were rewarded for making a Go response shortly after the presentation of a 35 dB sound pressure level (SPL) tone at any frequency. Training was performed in an acoustically transparent operant training chamber (20 x 20 x 18 cm, length x width x height) contained within a sound-attenuated chamber. Input and output devices (photobeam detector, food dispenser, sound card, house light) and software were manufactured by Med Associates (Georgia, VT). The rats were shaped in three phases. During phase A, rats were trained to make a nose poke response to obtain a food reward. During phase B, rats were trained to make a nose poke only after presentation of an auditory stimulus (5 kHz at 35 dB SPL). During phase C, rats were conditioned to make discriminative responses to the target stimulus and not to a limited set of non-target stimuli. Rats were advanced to the actual training program (levels 1–6) once their performance in phase C was clearly under stimulus control. Once rats were working within the actual training program, they began each day on level 1.

A single behavioral trial was defined as the length of time between the onsets of two successive tones. The intertrial interval was selected at random from a range of 3–9 s. A rat's behavioral state at any point in time could be classified as either Go or "NoGo." Rats were in the Go state when the photobeam was interrupted. All other states were considered NoGo. For a given trial, the rat could elicit one of five reinforcement states. The first four states were given by the combinations of responses (Go or NoGo) and stimulus properties (target or non-target). Go responses within 3 s of a target were scored as a hit, and a failure to respond within this time window was scored as a miss. A Go response within 3 s of a non-target stimulus was scored as a false positive, and the absence of a response was scored as a withhold. The fifth state, false alarm, was defined as a Go response that occurs >3 s after stimulus presentation. A hit triggered delivery of a food pellet (45 mg; BioServe, Frenchtown, NJ). A miss, false positive, and a false alarm initiated a 5 s "time-out" period during which time the house lights were turned off and no stimuli were presented. A withhold did not produce a reward or a time out.

Trials were grouped into blocks of 50. At the conclusion of each block, a target response ratio (number of hits/number of target trials), a non-target response ratio (number of false positives/number of non-target trials), and a false alarm ratio (number of false alarms/number of trials) were calculated. The target response ratio criterion was set to 80% and the non-target response ratio was set to 30%.

At the conclusion of training, each day of behavioral performance underwent an additional quality control. If the rat was inattentive (target response ratios <50%) or hyperactive (false alarm ratios >15%) over a block of 50 trials, the block was discarded. The rats worked for an average of 1200 acceptable trials per day. Less than 5% of the blocks met the criteria set for inattentive or hyperactive behavior.

## Surgery

Experiments were carried out in a sound-attenuating chamber. Each rat was anesthetized by intraperitoneal injection of sodium pentobarbital (50-80 mg/kg), with the dose adjusted to make the rat areflexic. The rat was maintained in an areflexic state for the rest of the experiment by further intraperitoneal injections of sodium pentobarbital (20-60 mg/kg) when necessary. The rat was placed on a heating pad, and its temperature maintained at ~37°C. Prior to any skin incision, bupivacaine was injected subcutaneously at the incision site. A tracheotomy was performed to secure the airway. The head was held fixed by a custom-made device that clamped it at both orbits and the palate, leaving the ears unobstructed. A cisternal drain was performed. The right auditory cortex was exposed by retracting the skin and muscle overlying it, followed by a craniotomy and a durotomy. The cortical surface was kept moist with normal saline. The auditory cortex was coarsely mapped at 500-600 μm below the pial surface with a parlyene-coated tungsten electrode to determine the location of intensity-tuned multiunit sites.

## Whole cell recordings

A silver wire, one end of which was coated with silver chloride, served as the reference electrode against which potentials were measured; its chlorided end was inserted between the skull and the dura. The reference electrode was assigned a potential of 0 mV. The potential of the cerebrospinal fluid was assumed to be uniform and equal to that of the reference electrode.

Patch pipettes with resistances of 7 MΩ were made. Patch pipettes contained a cesium based solution that consisted of (in mM) 130 Cs-gluconate, 5 TEA-Cl, 4 MgATP. 0.3 GTP, 10 phosphocreatine, 10 HEPES, 0.2 EGTA, 5 QX-314, pH 7.3. The pia was broken by slowly lowering and raising the jagged tip of a broken pipette in and out of the cortex. An unbroken pipette was lowered into the cortex, with the pressure in the pipette greater than atmospheric. Dimpling of the cortical surface was not visually detectable. The cortical depth of the pipette tip was estimated according to the distance it had travelled. The cortex was covered with 4% agarose in normal saline. An oscillating potential was set up at the pipette tip; the oscillating potential had a time average of −50 mV; its period was much faster than the breathing rate. The resulting current oscillation was measured. When the amplitude of the current oscillation decreased and an even slower oscillation whose period was the breathing rate of the animal became superimposed on the current oscillation, the pipette tip might be touching the cell membrane of a neuron. At this point, the pipette was slowly advanced to further reduce the amplitude of the current oscillation. The pressure in the pipette was suddenly reduced to less than atmospheric and then returned to atmospheric. Often a giga-ohm seal would spontaneously form within 1 min; if not, additional gentle suction sometimes helped. The pipette capacitance was compensated. A pulse of reduced pressure in the pipette would often break the cell membrane and bring the recording into whole cell mode. The whole cell capacitance was compensated and the initial series resistance (25-60 MΩ) was compensated to achieve an effective series resistance of 15-30 MΩ. The input resistance was 100-400 MΩ. Signals were filtered at 5 kHz and sampled at 10 kHz. A Multiclamp 700B amplifier was used.

**Stimuli**

Noise bursts were delivered by a calibrated free field speaker directed toward the left ear. The intensities of the noise bursts ranged from 0 dB SPL to 80 dB SPL, and were evenly spaced over that range. Noise bursts were white, 50 ms in duration with 5 ms linear rising and falling phases.

**Predicted membrane potential**

The predicted membrane potential $V_p$ was calculated using

$$C_m dV_p(t)/dt = G_r(t)(V_p(t)-E_r) + G_e(t)(V_p(t)-E_e) + G_i(t)(V_p(t)-E_i), \tag{1}$$

where $C_m$ is the membrane capacitance, $G_r$ the resting or leak conductance, $E_r$ the resting membrane potential, $G_e(t)$ the excitatory synaptic conductance, $G_i(t)$ the inhibitory synaptic conductance, and $E_e$ and $E_i$ are the reversal potentials of the excitatory and inhibitory synaptic conductances respectively. *Eq. 1* is a good model of the membrane voltage of cortical neurons in the absence of spiking (McCormick et al 1985; Troyer and Miller 1997). $C_m$ was derived from the whole cell capacitance compensation procedure. $G_r$ was derived using

$$I_r(V) = G_r(V-E_r), \tag{2}$$

where $V$ is the clamping voltage, and $I_r(V)$ the resting or leak current. Measurement of $I_r(V)$ at two different values of $V$ yielded a system of 2 equations which could then be solved for $G_r$. $E_r$ was measured in current clamp. $G_e(t)$ and $G_i(t)$ were derived using

$$\Delta I(t,V) = G_e(t)(V-E_e) + G_i(t)(V-E_i), \tag{3}$$

where $V$ is the clamping voltage, $\Delta I(t,V)$ the amplitude of the synaptic current, relative to the resting current at $V$. The values of $E_e$ and $E_i$ were set by the ionic composition of the pipette solution and the cerebrospinal fluid (Johnston and Wu 1995; Davson and Segal 1996); the value of $E_i$ was based on the permeability of GABA-A conductances to $Cl^-$, but it should be noted that they also pass $HCO_3^-$ (Bormann et al 1987). $G_e(t)$ and $G_i(t)$ were the 2 unknowns in *Eq. 3* at any particular $t$. Measurement of $\Delta I(t,V)$ at two different values of $V$ yielded a system of 2 equations which could then be solved for $G_e(t)$ and $G_i(t)$ at any particular $t$ (Borg-Graham et al. 1998; Anderson et al. 2000; Hirsch et al. 1998). Currents into the neuron were assigned a negative value. $E_e$ and $E_i$ were 0 mV and -70 mV respectively. We did not correct the measured membrane potential for the series resistance or the junction potential.

## Results

Intensity-tuned neurons are not common in the rat auditory cortex. To increase the chances of obtaining in vivo whole cell recordings from intensity-tuned neurons, rats underwent training that increases the proportion of intensity-tuned neurons. Rats were anesthetized with pentobarbital and their auditory cortices coarsely mapped to locate intensity-tuned multiunit sites, to which whole cell recording attempts were then directed. Noise bursts ranging from 0 dB to 80 dB in 5 dB steps, with at least 10 repetitions per intensity, were used to determine intensity-tuning. The patch pipette contained cesium, TEA, and QX-314, which blocked spiking and most intrinsic conductances, and prevented them from contaminating measurements of the synaptic conductances. Because spiking was blocked, the peak membrane potential was used as gauge of spike rate. Synaptic excitation was seen as inward currents when the neuron was voltage-clamped at -70 mV, near the inhibitory reversal potential; synaptic inhibition was seen as outward currents when the neuron was clamped more positively, nearer the excitatory reversal potential. Whole cell recordings were obtained from 17 neurons at intensity-tuned multiunit sites. Unbalanced synaptic excitation and inhibition were demonstrated in 5 neurons. These 5 neurons are described below.

Responses from an unbalanced neuron are shown in Fig. 1. Fig. 1A shows the average synaptic currents obtained at 45 dB and 70 dB. The inward currents representing synaptic excitation (blue traces, downward deflections) were obtained when the neuron was clamped at -70 mV, and the outward currents representing synaptic inhibition (red traces, upward deflections) when the neuron was clamped at 0 mV. The synaptic excitation at both intensities has the same amplitude, but no synaptic inhibition is observed at 45 dB, while there is substantial synaptic inhibition at 70 dB. This suggests that synaptic inhibition would cause less depolarization of the membrane potential at 70 dB than at 45 dB, and thus contribute to some intensity tuning in this neuron. Fig. 1B shows that synaptic excitation itself is nonmonotonic, showing a distinct peak around 40 dB to 50 dB. From 60 dB to 70 dB, inhibition increases faster than excitation, suggesting that inhibition enhances the intensity-tuning of this neuron. However, from 75 dB to 80 dB, inhibition decreases as excitation increases. Thus this neuron might show 2 peaks in its intensity-tuning profile. It is also possible that the timing of the inhibition at 75 dB and 80 dB is shifted, so as to decrease the membrane depolarization at those intensities. Distinguishing these possibilities requires a record of the membrane potential, which was not obtained for this neuron.

Responses from a second, unbalanced intensity-tuned neuron are shown in Fig. 2. Fig. 2A shows the average membrane potential responses at 15 dB, 25 dB and 75 dB. The noise-evoked depolarization at 25 dB is greater than at 15 dB, but that at 75 dB is less than at 25 dB. Fig. 2B graphs the peak average membrane potential versus sound intensity. The peak average membrane potential increases to a maximum at 25 dB, then decreases above that, showing that this neuron is intensity-tuned. (Figs. 2C and 2D show the same data as Figs. 2A and 2B respectively, but with the membrane potential immediately preceding sound onset subtracted from each trace before averaging.) Fig. 2E shows the average synaptic currents obtained at 25 dB and 80 dB. When the neuron is clamped at -70 mV the inward currents (blue traces, downward deflections) representing synaptic excitation are equal for 25 dB and 80 dB. In contrast, at -30 mV the outward current (red traces, upward deflections) representing synaptic inhibition is much greater at 80 dB than at 25 dB. The synaptic current at -30 mV at 25 dB displays an initial inward portion that represents synaptic excitation. This inward current at -30 mV is as large as the inward current at -70 mV, indicating that the excitatory conductance has a nonlinear dependence (possibly the NMDA channel) on the membrane potential. This nonlinearity might cause synaptic inhibition to be underestimated. It will not, however, affect the conclusion that synaptic excitation and inhibition are unbalanced in this neuron. This is because the inward current at -70 mV is the same at both 25 dB and 80 dB, indicating that the excitatory conductance is the same at both intensities. Thus, the nonlinearity observed in the synaptic current at -30 mV should be present in equal amounts for both 25 dB and 80 dB. Unlike the synaptic current at -30 mV at 25 dB, there is almost no indication of the nonlinearity in the synaptic current at -30 mV at 80 dB because the synaptic current is disproportionally very much larger. Fig. 2F graphs the peak average inward and outward currents (blue and red lines respectively) versus intensity. The magnitude of the peak average inward current reaches a maximum at 25 dB, then decreases above that. The maximum of the peak average inward current occurs at the same intensity as that of the peak average membrane potential, showing that the intensity-tuning of the membrane potential is partially present in the excitation. Yet the peak average inward current increases above 65 dB, while the peak average membrane potential decreases, showing that excitation alone cannot account for the membrane potential at intensities above 65 dB. In that intensity range, the increase with intensity of the peak average outward current is much sharper than that of the peak average inward current. This increasing ratio of inhibition to excitation can explain the continued decrease in membrane potential above 65 dB. To confirm these points, we used the excitatory and inhibitory currents to predict the membrane potential. Fig. 2G shows example predicted membrane potential traces at 15 dB, 25 dB and 75 dB; these resemble the actual membrane potential traces at those intensities in Fig. 2A. The graph of peak predicted membrane potential versus intensity shown in Fig. 2H matches the actual curve in Fig. 2B. This neuron thus provides an example of an unbalanced intensity-tuned neuron in which synaptic inhibition enhances, but does not create intensity-tuning.

Responses from a third, unbalanced intensity-tuned neuron are shown in Fig. 3. Fig. 3A shows the average membrane potential responses at 3 intensities. The noise-evoked depolarization at 25 dB is greater than at 55 dB, but that at 80 dB is less than at 55 dB. Fig. 3B graphs the peak average membrane potential versus sound intensity. The peak average membrane potential increases to a maximum at 55 dB, then decreases above that, showing that this neuron is intensity-tuned. Fig. 3E shows the average synaptic currents obtained at 55 dB and 80 dB. At -70 mV, the inward currents representing synaptic excitation are equal for 55 dB and 80 dB. In contrast, at -20 mV the outward current representing synaptic inhibition is much greater at 80 dB than at 25 dB. Like the neuron of Fig. 2, there is thus disproportionally large inhibition at the higher intensity which will either enhance or create intensity-tuning. Fig. 3F graphs the peak average inward and outward currents versus intensity. Unlike the neuron of Fig. 2, the peak average inward current is not intensity-tuned, but increases monotonically with intensity. Therefore the intensity-tuning must be created by synaptic inhibition. The sharper increase with intensity of the peak average outward current over that of the peak average inward current for intensities above 55 dB, produces an increasing ratio of inhibition to excitation that can explain the continued decrease in membrane potential above 55 dB. Fig. 3G shows example predicted membrane potential traces at 25 dB, 55 dB and 80 dB; these resemble the actual membrane potential traces at those intensities in Fig. 3A. The graph of peak predicted membrane potential versus intensity shown in Fig. 3H matches the actual curve in Fig. 3B. This neuron thus provides an example of an unbalanced intensity-tuned neuron in which synaptic inhibition actually creates intensity-tuning.

The responses from a fourth, unbalanced intensity-tuned neuron are shown in Fig. 4. Again, Figs. 4A and 4B demonstrate that the membrane potential is intensity-tuned. (Figs. 4C and 4D show the same data as Figs. 4A and 4B respectively, but with the membrane potential immediately preceding sound onset subtracted from each trace before averaging. In this case, examining the peak membrane potential suggests, but does not confirm intensity-tuning: the curves of Figs. 4B and 4D have maxima at different intensities, indicating that the intensity at which the maximum occurs is uncertain, and suggesting that the intensity at which the minimum occurs is also uncertain. However, Figs. 4A and 4C show that the membrane potential at 80 dB, compared with that at 60 dB, clearly exhibits only the shortest depolarization, but a sustained and much greater hyperpolarization, confirming that the membrane potential is intensity-tuned.) Fig. 4E shows examples of the excitatory and inhibitory synaptic currents, showing disproportionally large inhibition at the higher intensities. Fig. 4F shows that the peak average excitatory currents are monotonic, and display no intensity-tuning. Therefore the intensity-tuning must be created by synaptic inhibition. The peak average inhibitory current does indeed increase more quickly from 65 dB to 70 dB. However, it increases just as quickly as the excitation from 70 dB to 80 dB, even though the membrane potential continues to decrease in that range (Fig. 4B). A linear least squares fit of the peak of inhibition versus the peak of excitation yields a correlation coefficient whose square is 0.63, within the range of a balanced neuron (Wehr and Zador 2003; Zhang et al 2003). This suggests that changes in the timing of the inhibition relative to the excitation are responsible for the further decrease of the membrane potential from 70 dB to 80 dB. To test this, we used the excitatory and inhibitory currents to predict the membrane potential. Fig. 4G shows example predicted membrane potential traces at 60 dB and 80 dB; these resemble the actual membrane potential traces at those intensities in Fig. 4A. The graph of peak predicted membrane potential versus intensity is shown in Fig. 4H, and resembles the actual curve of Fig. 4B. The final fall-off in the peak predicted membrane potential is too small to indicate intensity-tuning. As was the case with the peak membrane potential, examining the peak predicted membrane underestimates intensity-tuning. Fig. 4G shows that the predicted membrane potential at 80 dB, compared with that at 60 dB, clearly exhibits only the shortest depolarization, but a sustained and much greater hyperpolarization, confirming that the predicted membrane potential is intensity-tuned.

Fig. 5 shows a fifth, unbalanced intensity-tuned neuron. Once again, Figs. 5A and 5B demonstrate that the membrane potential is intensity-tuned. Fig. 5E shows examples of the excitatory and inhibitory synaptic currents. Fig. 5F shows that the synaptic excitation is monotonic. Thus synaptic inhibition must create the intensity tuning. However, a linear least squares fit of the peak of inhibition versus the peak of excitation yields a correlation coefficient whose square is 0.90, well within the range of a balanced neuron (Wehr and Zador 2003; Zhang et al 2003). It appears that changes in the relative timing of excitation and inhibition account for the intensity-tuning of the membrane potential. This is confirmed by using the synaptic currents to predict the membrane potential. Fig. 5G shows example predicted membrane potential traces at 3 intensities; these resemble the actual membrane potential traces at those intensities in Fig. 5A. The graph of peak predicted membrane potential versus intensity shown in Fig. 5H matches the actual curve in Fig. 5B.

## Discussion

We have described 5 unbalanced intensity-tuned neurons in the auditory cortex. In some of these synaptic inhibition enhanced the intensity-tuning; in others it even created the intensity-tuning. Since synaptic inhibition is essentially cortical in origin, the latter have provided examples of auditory feature-selectivity arising de novo at the auditory cortex.

We used the peak membrane potential as a gauge of spike rate; the precise relationship between them requires further investigation. Entering the measured synaptic currents into a model Hodgkin-Huxley neuron might give a good estimate of the spike responses of these neurons. A simple approximation to the spike rate generated by a Hodgkin-Huxley neuron is the net synaptic current at the spike threshold; however, it is uncertain if this holds for transient synaptic inputs like those observed here (Koch et al 1995, Wilent and Contreras 2005). Experimentally, this may be addressed by recording the spikes in cell-attached mode before the whole-cell mode is established, or by removing pharmacological blockers of spiking from the intracellular solution (Borg-Graham et al 1998, Wehr and Zador 2003).

Intensity-tuned neurons are generated by diverse patterns of synaptic inhibition, from the balanced intensity-tuned neurons observed by Wehr and Zador (2003), to the several patterns of unbalanced synaptic excitation and inhibition described here. Perhaps these different patterns occur in different layers or areas of the auditory cortex. In the cat primary auditory cortex, the distributions of inhibitory neurons and inhibitory axon terminals depend on layer and position in the isofrequency contour (Hendry and Jones 1991; Prieto et al 1994a, 1994b). Tones evoke greater synaptic inhibition in layer 3 neurons than in layer 2 neurons (Ojima and Murakami 2002). Diverse patterns of synaptic excitation and inhibition are also present in the primary visual cortex, with some of the patterns organized by layer (Borg-Graham et al 1998, Anderson et al 2001; Martinez et al 2002; Monier et al 2003).

A previous study addressed the role of synaptic inhibition in creating intensity-tuning by using a pharmacological blocker of synaptic inhibition. They found that only 2 of 31 intensity-tuned neurons became monotonic when synaptic inhibition was blocked.

However, they also found that 12 of 38 monotonic neurons became intensity tuned when inhibition was blocked. It is not straightforward to fit their data and ours into a single picture. Disrupting inhibition pharmacologically affects the inhibition at all neurons in a network in which there may be complex feedback connections; for a straightforward comparison with our work, inhibition should be disrupted at only a single neuron. Disrupting inhibition pharmacologically also raises the spike rate of all cells, which may produce intensity-tuning via a combination of synaptic and voltage-gated intrinsic conductances (Sivaramakrishnan et al 2004); in comparison we have examined only synaptic conductances.

As behavioural training can convert balanced, monotonic neurons into intensity-tuned neurons, the question is raised as to whether the new intensity-tuned neurons are balanced or unbalanced, and if they are unbalanced, whether inhibition enhances or creates their intensity-tuning. Extracellular recordings show that training doubles the number of intensity-tuned neurons (Polley et al 2004). If an overwhelming majority of intensity-tuned neurons in trained animals are unbalanced, that would suggest that balanced, monotonic neurons are being converted to unbalanced, intensity-tuned neurons. This would place a severe constraint on models of cortical circuitry, which would have to robustly ensure that monotonic neurons in an untrained animal are balanced, but not so robustly that training cannot undo it. Unfortunately, the small number of neurons in our sample and the heterogeneity of the patterns of synaptic excitation and inhibition observed make it impossible, at the moment, for us to determine the ratio of balanced to unbalanced intensity-tuned neurons in trained animals.

## Acknowledgements

We thank Julie L. Schnapf and Michael P. Stryker for their comments on a draft of this paper. This work was supported by the National Institutes of Health.

FIG. 1. Unbalanced neuron. A: Average synaptic currents at 2 intensities, at -70 mV (blue) or 0 mV (red). B: Peak average inward (blue) and outward (red) currents versus sound intensity.

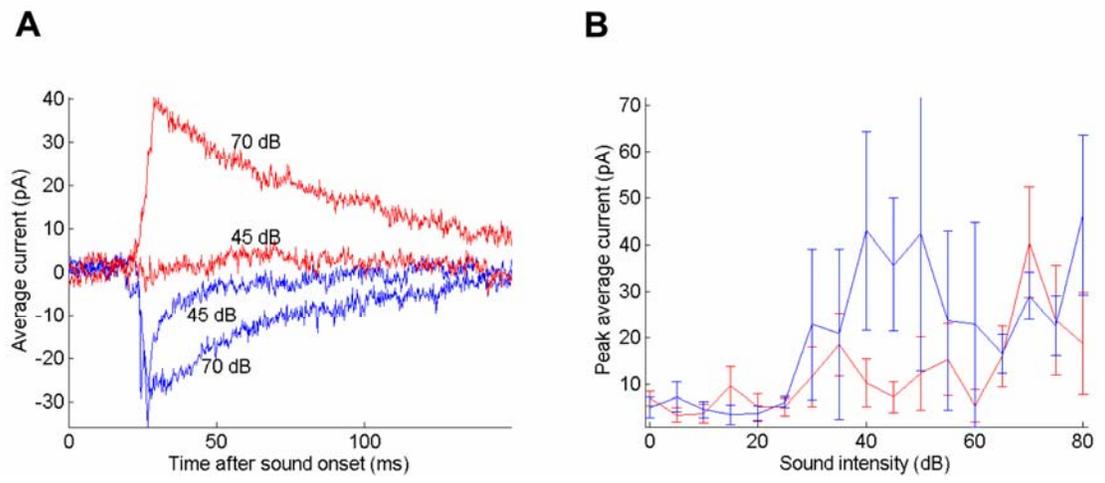

Figure 1

FIG. 2. Unbalanced neuron in which inhibition enhances intensity-tuning. A: Average membrane potential responses at 3 intensities. B: Peak average membrane potential versus sound intensity. C: Same as A, but with the membrane potential immediately preceding sound onset subtracted from each trace before averaging. D: Same as B, but with the membrane potential immediately preceding sound onset subtracted from each trace before averaging. E: Average synaptic currents at 2 intensities, at -70 mV (blue) or -30 mV (red). F: Peak average inward (blue) and outward (red) currents versus sound intensity. G: Predicted membrane potential responses at 3 intensities. H: Peak predicted membrane potential versus sound intensity.

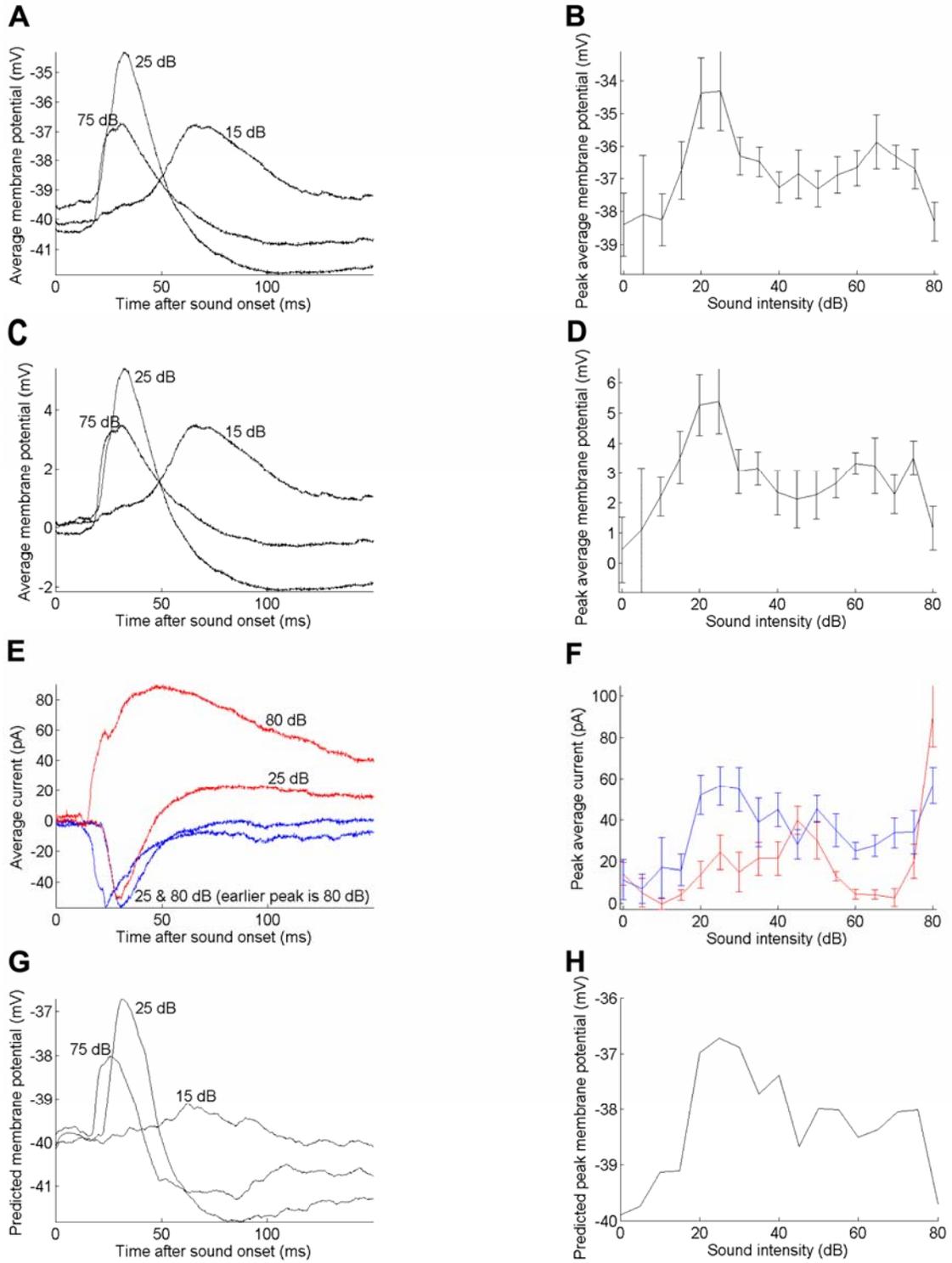

Figure 2

FIG. 3. Unbalanced neuron in which inhibition creates intensity-tuning. A: Average membrane potential responses at 3 intensities. B: Peak average membrane potential versus sound intensity. C: Same as A, but with the membrane potential immediately preceding sound onset subtracted from each trace before averaging. D: Same as B, but with the membrane potential immediately preceding sound onset subtracted from each trace before averaging. E: Average synaptic currents at 2 intensities, at -70 mV (blue) or -20 mV (red). F: Peak average inward (blue) and outward (red) currents versus sound intensity. G: Predicted membrane potential responses at 3 intensities. H: Peak predicted membrane potential versus sound intensity.

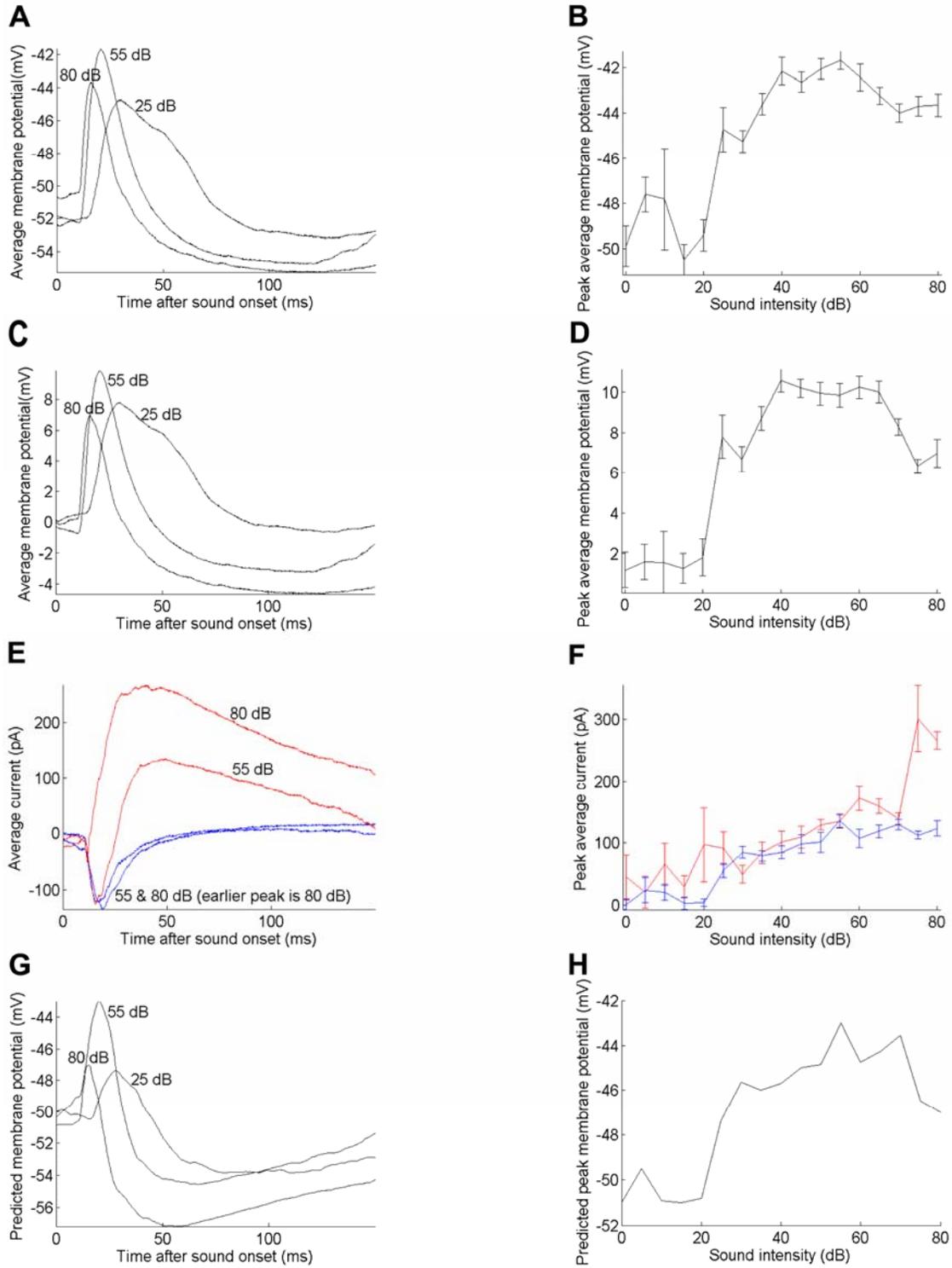

Figure 3

FIG. 4.   Unbalanced neuron in which inhibition creates intensity-tuning. A: Average membrane potential responses at 3 intensities. B: Peak average membrane potential versus sound intensity. C: Same as A, but with the membrane potential immediately preceding sound onset subtracted from each trace before averaging. D:  Same as B, but with the membrane potential immediately preceding sound onset subtracted from each trace before averaging. E: Average synaptic currents at 2 intensities, at -70 mV (blue) or -30 mV (red). F: Peak average inward (blue) and outward (red) currents versus sound intensity. G: Predicted membrane potential responses at 2 intensities. H: Peak predicted membrane potential versus sound intensity.

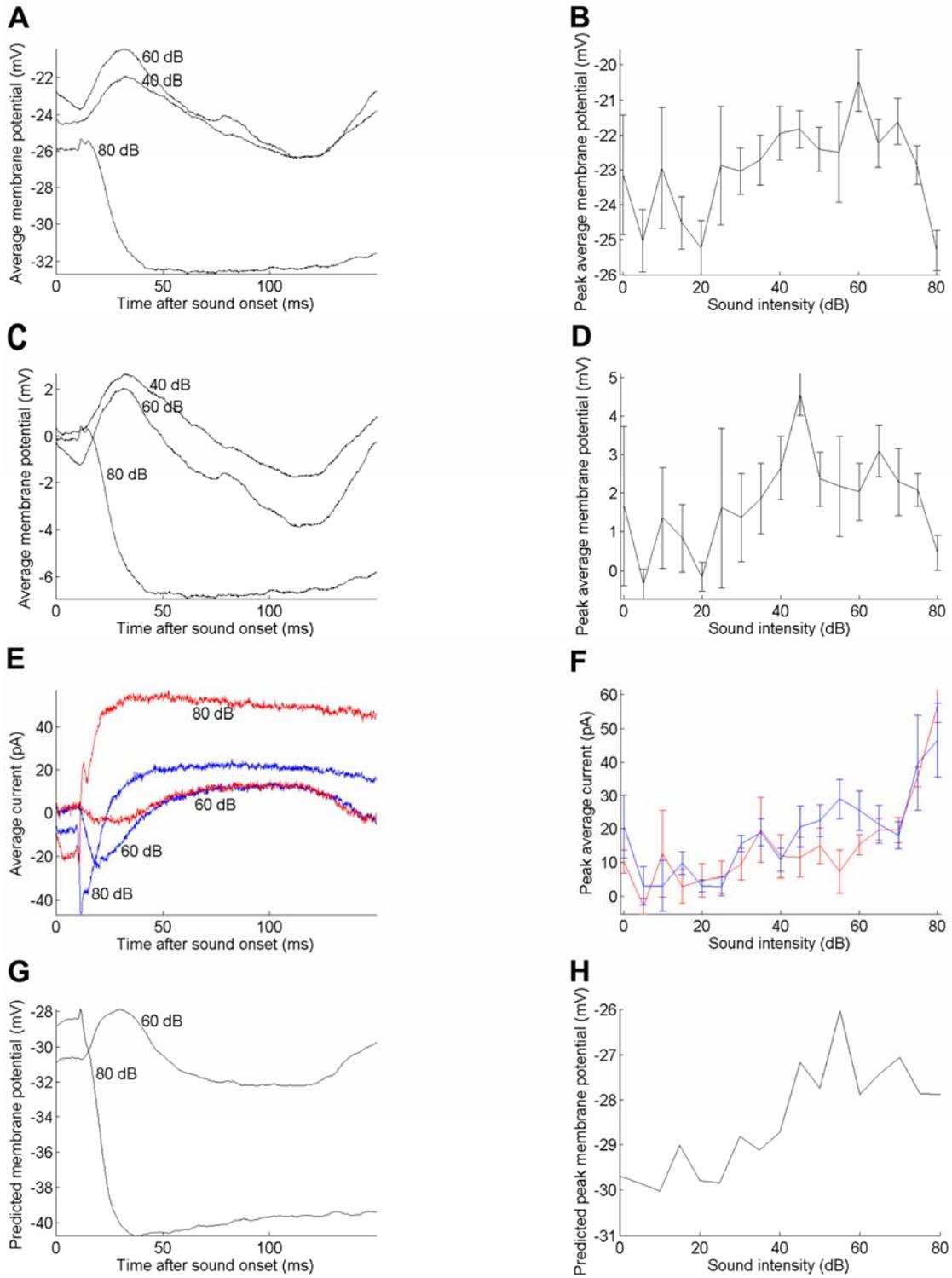

Figure 4

FIG. 5. Unbalanced neuron in which inhibition creates intensity-tuning. A: Average membrane potential responses at 3 intensities. B: Peak average membrane potential versus sound intensity. C: Same as A, but with the membrane potential immediately preceding sound onset subtracted from each trace before averaging. D: Same as B, but with the membrane potential immediately preceding sound onset subtracted from each trace before averaging. E: Average synaptic currents at 2 intensities, at -70 mV (blue) or 0 mV (red). F: Peak average inward (blue) and outward (red) currents versus sound intensity. G: Predicted membrane potential responses at 2 intensities. H: Peak predicted membrane potential versus sound intensity.

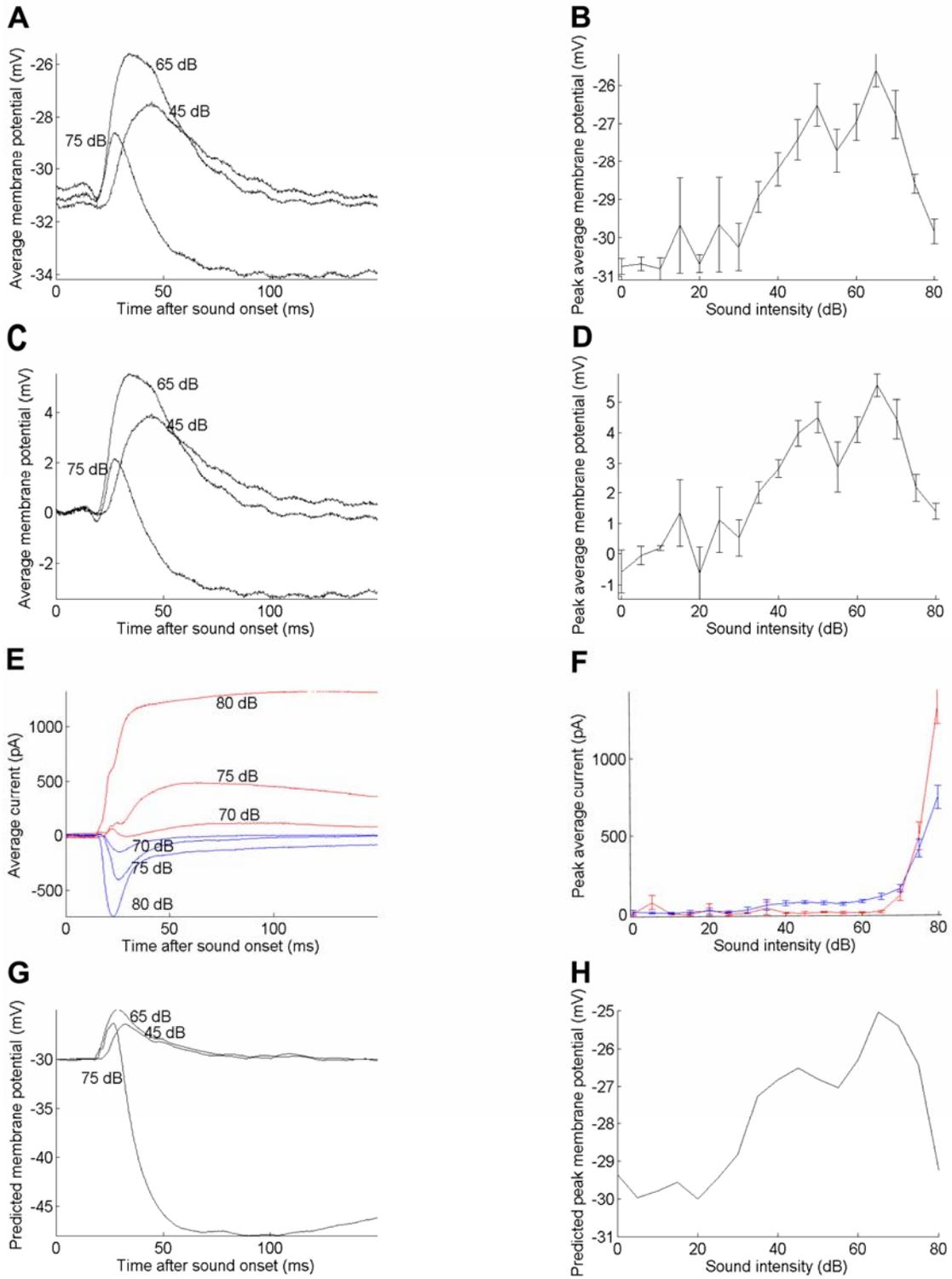

Figure 5